# AN INTELLIGENT CLASSIFICATION MODEL FOR PHISHING EMAIL DETECTION

Adwan Yasin and Abdelmunem Abuhasan

College of Engineering and Information Technology, Arab American University, Palestine

*ABSTRACT*

*Phishing attacks are one of the trending cyber-attacks that apply socially engineered messages that are communicated to people from professional hackers aiming at fooling users to reveal their sensitive information, the most popular communication channel to those messages is through users' emails. This paper presents an intelligent classification model for detecting phishing emails using knowledge discovery, data mining and text processing techniques. This paper introduces the concept of phishing terms weighting which evaluates the weight of phishing terms in each email. The pre-processing phase is enhanced by applying text stemming and WordNet ontology to enrich the model with word synonyms. The model applied the knowledge discovery procedures using five popular classification algorithms and achieved a notable enhancement in classification accuracy; 99.1% accuracy was achieved using the Random Forest algorithm and 98.4% using J48, which is –to our knowledge- the highest accuracy rate for an accredited data set. This paper also presents a comparative study with similar proposed classification techniques.*

*KEYWORDS*

*phishing, data mining, email classification, Random Forest, J48.*

## 1. INTRODUCTION

The recent advances in web and mobile technology attracted most commercial institutions to offer their services online, including banks, stocks and ecommerce providers. As people increasingly rely on Internet services to carry out their transactions, Internet fraud becomes a great threat to people's privacy and safety. Phishing is one of the main types of Internet fraud; which relies on fooling users to share or declare their private information (including passwords and credit card numbers), phishing could be defined as a cyber-attack that communicates socially-engineered messages to humans through electronic communication channels (email, SMS, phone call) in order to persuade them to do certain actions (enter credentials, credit card number, …) for the attackers benefit; such actions could be persuading an e-commerce web site user to enter his credentials to a fake web site (managed by the attacker) similar to the original website and then the attacker uses them to impersonate the user. In order to persuade the victim user to login to such a fake website, the socially engineered message draws an illusion to the user that he needs to perform such action, such as warning the user about account suspension or that the website admin is requesting him to reset his password [1].

Phishing attacks employ email messages and websites that are designed in a professional manner to be similar to emails and websites from legitimate institutions and organizations (usually the user is a customer for those organizations), to persuade users into disclosing their personal or financial information. The attacker can then use collected sensitive user information for his benefit. Users can be tricked into disclosing their information either by providing sensitive information via a web form, replying to spoofed emails, or downloading and installing Trojans, which search users' computers or monitor users' online activities in order to get information.

Phishing attacks have steadily increased to match the growth of electronic commerce, recently taking on epidemic proportions; the Anti Phishing Work Group (APWG) report of 2015 [2] declared that the total number of unique phishing sites detected from Quarter1 through Quarter3 of 2015 was 630,494, while The number of unique phishing reports submitted to APWG from quarter 1 through quarter3 was 1,033,698. According to a recent study from Google [3], 45% of phishing websites fooled their target victims into declaring their passwords, and got their password changed by the attacker within 30 minutes after their accounts were hijacked. The attackers also exploited the victims' accounts in fooling other people in the victim's contact list through communicating with them using the hijacked accounts; the study concluded that those people are 36 times more likely to be hijacked when the attackers used the victim's account to communicate with them, and this is an expected result as the communication is received from a trusted account.

Many researchers have studied the phishing problem and proposed a variety of solutions to combat phishing attacks. The first category of proposed solutions works on the principle of detecting phishing attacks and warning the user or preventing him from taking actions that could result in compromising his private data, latest research proposals in this category include [4] [5] [6] [7] [8] [9]. The second category of proposed phishing solutions rely on securing the login process by adding a second authentication factor such that stealing the user's credentials is not enough for an attacker to compromise the victim's account unless he also possesses the second authentication factor, those proposals include [10] [11] [12] [13] [14] [15] [16] [17] [18] [19].

Our focus in this paper is to build an intelligent classifier at the email level that is capable of detecting phishing emails as an early stage in the phishing combating process; we believe that detecting phishing emails can make the internet users more secure by eliminating those emails and not relying on the users' vigilance to protect them from phishing attacks; many studies concluded that depending on human factors is not a preferred option for combating phishing attacks; especially for advanced and well prepared phishing attacks that are continuously adapting themselves to known defence mechanisms [20] [21].

Our approach for detecting phishing emails applies the knowledge discovery model and data mining techniques to build an intelligent model that learns from existing training dataset of both ham and phishing emails, the model will extract and reduce the important features that contribute to building a set of classifiers from which the best classifier is chosen. We built a java program that extracts a set of features from the email header and body, those features are then augmented with a weighted term frequency that is applied after performing linguistic processing of the email extracted terms. After that a set of data mining algorithms are applied to the extracted features to decide the algorithm with best results.

## 2. RELATED WORK

In the study [8] the authors proposed a model that utilizes 23 hybrid features of the email header and body extracted from about 10000 emails divided equally between ham and spam emails, their model applied J48 classification algorithm to classify phishing and legitimate emails and concluded with an accuracy of 98.11% and false positive rate of 0.53%.

Another study [22] applied a two-phase classification model of emails; in the first phase a set of classification algorithms (C5.0, Naive Bayes, SVM, Linear Regression and K-Nearest Neighbours) are used to classify legitimate and phishing emails, common evaluation metrics are used to evaluate each algorithm including accuracy, precision, recall and F-score, the algorithm with best classification results was C5.0 with an average accuracy rate of 97.15%, average precision of 98.56%, average recall of 95.64% and average F-score of 97.08%. in the second phase, the emails that were classified as legitimate in the first phase were input to an ensemble classifier.

The authors in [23] proposed an email classification model that exploits 23 keywords extracted from the email body, the proposed model was tested using a set of classification algorithms, including multilayer perceptron, decision trees, support sector machine, probabilistic neural net, genetic programming, and logistic regression. The best classification result was achieved using genetic programming with a classification accuracy of 98.12%.

The study [24] applies the Bayesian classifier for phishing email detection, evaluated in terms of accuracy, error, time, precision and recall. The model resulted in accuracy of 96.46%.

The authors in [25] applied Support Vector Machine classifier to classify emails using a set of 9 structure-based and behaviour-based features. The model achieved 97.25% accuracy in results, however its weakness is in its relatively small training dataset (1000 emails with 50% spam and 50% ham).

The authors in [26] proposed an email classification algorithm by integrating Bayesian Classifier and phishing URLs detection using Decision Tree C4.5, their approach achieved 95.54 % accuracy, which is better than the accuracy of 94.86% that was achieved using Bayesian classifier.

The study in [27] used Random Forest and Partial Decision Tree algorithm for spam email classification, the authors applied a set of feature selection methods in the pre-processing step including Chi-square and Information gain, they achieved accuracy of 96.181% with Random Forest and 95.093% with Part.

The authors in [28] proposed a browser knowledge-based compound approach for detecting phishing attacks, the proposed model analyses web URLs using parsing and utilizes a set of maintained knowledge bases which store the previously visited URLs and previously detected phishing URLs. The experimental results indicated 96.94% accuracy in detecting phishing URLs with a little compromise in degrading the browser speed.

## 3. PROPOSED MODEL

The proposed approach for phishing email classification employs the model of Knowledge Discovery (KD) and data mining for building an intelligent email classifier that is able to classify a new email message as a legitimate or spam; the proposed model is built by applying the iterative steps of KD to identify and extract useful features from a training email data set, the features are then fed to a group of data mining algorithms to identify the best classifier.

The proposed model for email classification utilizes linguistic processing techniques and ontologies to enhance the similarity between emails with similar semantic term meaning, also the principle of term document frequency is applied in weighting the phishing terms in each email such that emails phishing terms weighting helps in discriminating phishing from legitimate emails. The proposed model also reduced the number of features used in the classification process into 16 features only; which enhances the classification performance and efficiency and minimizes the noise of including many features and hence improves the classification accuracy. These enhancements and are discussed in detail in the following subsections.

### 3.1 Knowledge Discovery Model

Knowledge discovery is the process of extracting or discovering patterns from data, the extracted patterns should be novel, valid, useful and understandable [29]. The KD process is carried out using a set of iterative steps as depicted in figure 1. The steps are initiated by understanding the problem and the data, followed by a data pre-processing phase to prepare it for the data mining step through which the target knowledge is discovered, evaluated and then presented as a useful and easy to use information.

The proposed model architecture is depicted in figure 2 and explained subsequently.

## 3.2 Data Collection

The first step in building the proposed phishing email classifier is choosing the suitable training data set which is a real sample of existing emails that consists of both phishing and legitimate emails (also known as spam and ham emails). The training data set will be used to discover potentially predictive relationships that will serve as building blocks in the classifier. Our training data set consists of 10538 emails including 5940 ham emails from spam assassin project [30] and 4598 spam emails from Nazario phishing corpus [31].

## 3.3 Data Pre-processing and features extraction

In this step the emails in the training data set are prepared and filtered such that they can be transformed into a data format that is easily and effectively processed in subsequent steps of building the classifier. The emails in our chosen training data set are available in plain text format which needs to be pre-processed and transformed into EML format (Microsoft Outlook Express file extension) that is interoperable with the java mail package that will be used to extract the email features. Figure 3 depicts the main actions that take place in the pre-processing step.

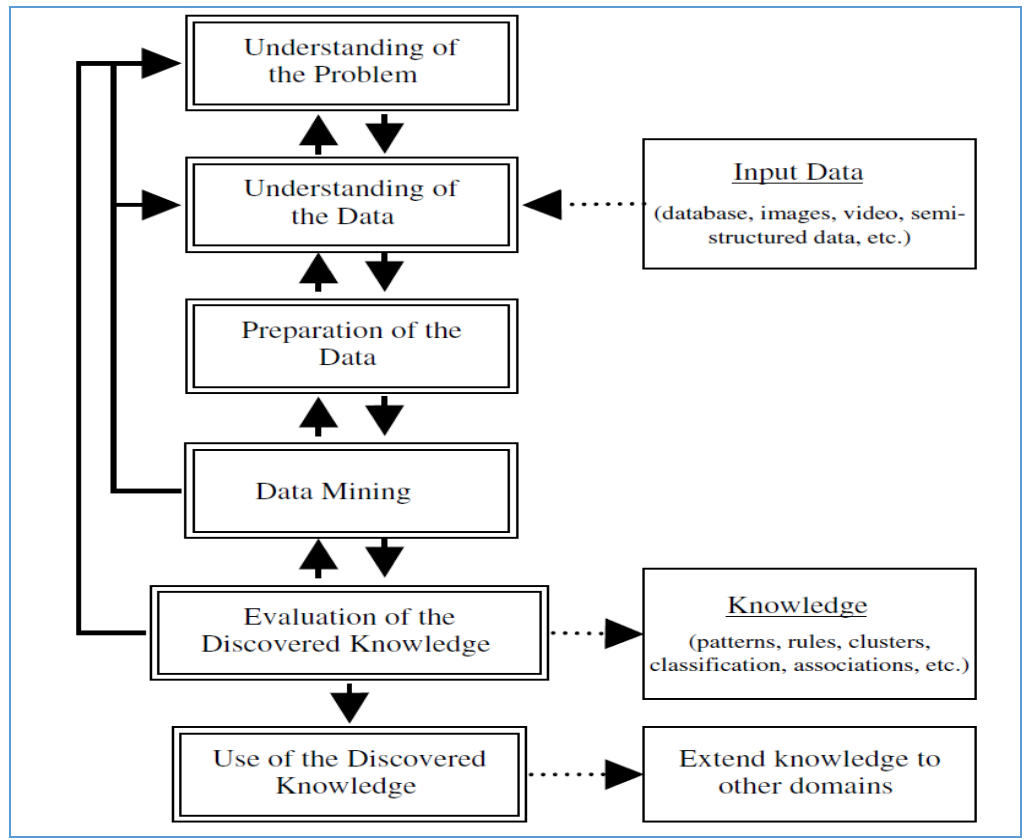

Figure 1: Knowledge Discovery Process [32]

The proposed mail classification model utilizes a set of 16 extracted features from the email message header and body, the extracted features are explained in table 1.

The process of extracting the features set from each email utilizes a java program that reads each email in the training data set, parses its contents and computes the value for each feature according to its description, after extracting the feature set for each email it is written into an ARFF (Attribute-Relation File Format) file that will be fed later into the classifier building process.

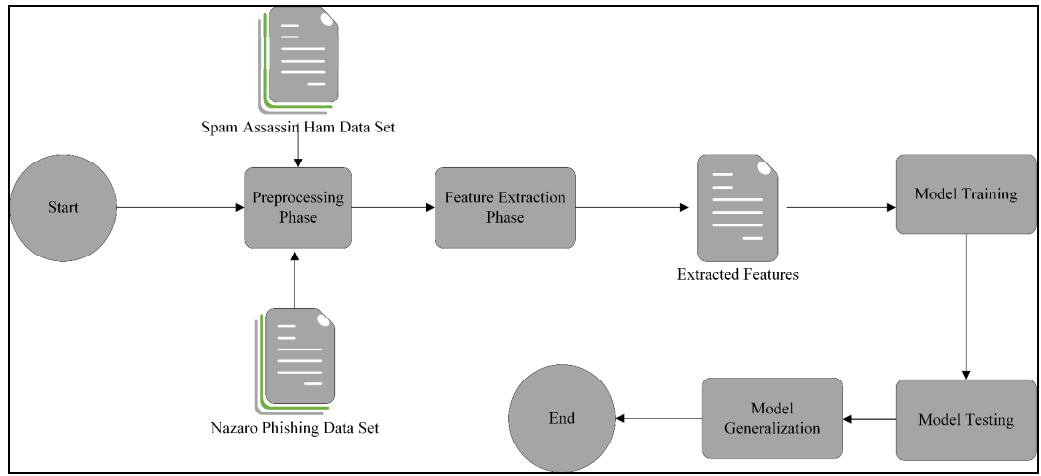

Figure 2: The proposed model architecture

We used the Information Gain (IG) measure to specify the usefulness of each feature in our features set in discriminating between the spam and ham classes, the IG value for each attribute tells us how important a given feature of the feature vectors is. The IG for each feature is depicted in table1. We found that the features listed in table 1 has the highest IG value which indicates that they will have an important contribution in deciding the email class as phishing or legitimate.

Figure 4 depicts the IG values of our proposed feature set.

The pre-processing phase consists of a set of steps that utilizes the email header, body and text features to extract the features that contribute to the classification process, some features are extracted from the URL links in the email subject and body, such as Hexadecimal URLs, Domains Count, TextLinkDifference, Dots Count, Images as URL and IP URLs. Other features are extracted from the email body such as HTML Body feature, the rest of features are extracted after processing the email subject and body text, this text processing step includes the following tasks:
- **Text parsing, tokenization and stemming**: the email subject and body text is parsed and tokenized into tokens, if the email body is HTML-formatted then the HTML tags are parsed to extract the text and identify URLs. Moreover, if the email contains attachments, they will also be parsed and tokenized. Each token in the extracted token is normalized such that morphological and in flexional endings of the tokens are removed, this stemming process is carried out using Porter Stemmer [33].
- **Stop words removal**: in this step, extremely common words which would appear to be of little value are removed from the extracted tokens, common stop words include the tokens "the","then","he",…etc. this step helps in reducing the similarities between emails and improves the performance of the proposed model specially in executing later steps.

International Journal of Network Security & Its Applications (IJNSA) Vol.8, No.4, July 2016

- **Semantic text processing:** in this step, each token in the email is augmented with its conceptually-related words from the WordNet ontology [34] using the synonymy and hyponymy relationships, this step helps identifying semantic relationships between tokens in different email messages and thus shortening the distance between feature vectors that contain close proximity to one another, and hence enhances the classification accuracy.

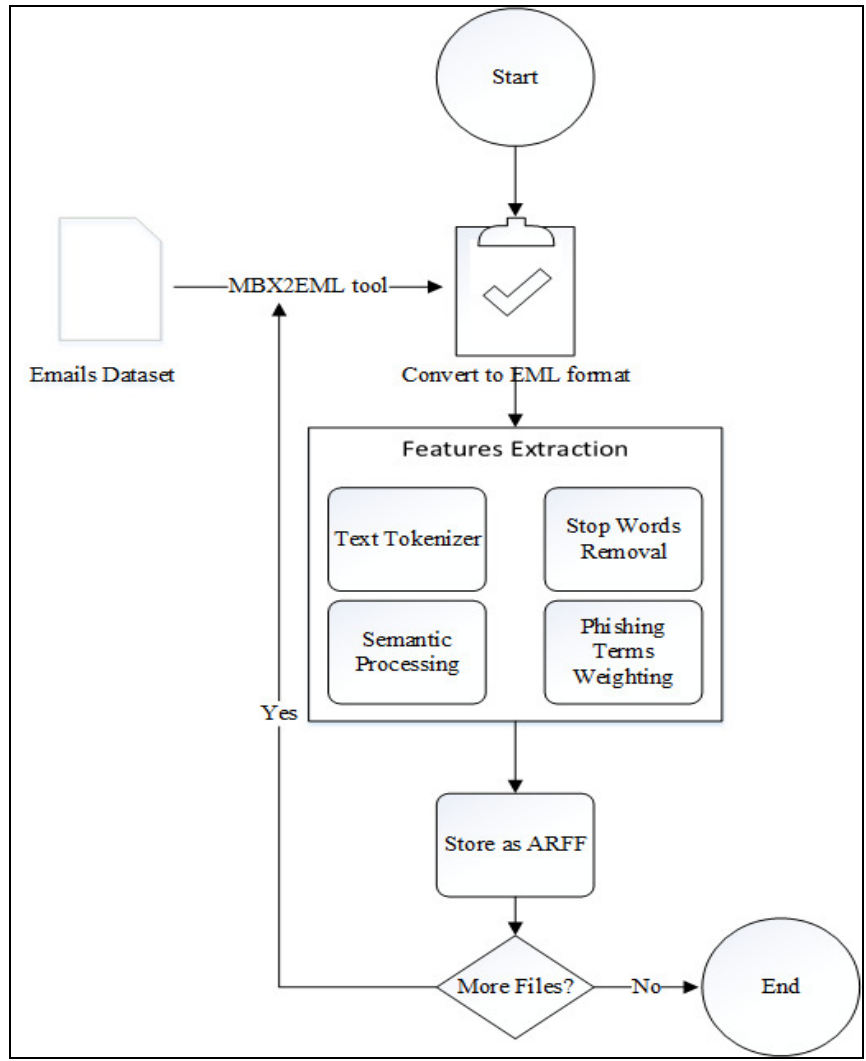

Figure 3: Pre-processing Phase

- **Phishing terms weighting**: in this step a set of phishing terms is built using the phishing emails in the training dataset. The phishing terms are those who have highest term frequency in the phishing data set. For example, the terms "Account" and "Please" existed in the phishing corpus 3384 and 3149 times respectively. This high frequency of terms indicates their importance in identifying phishing emails.

The proposed preprocessing model extracts the set of phishing terms- denoted by PT- from the set of phishing emails in the training data set, the phishing terms should also be not included in the legitimate emails training data set. The PT data set includes all terms whose document frequency (the number of phishing emails that contain the phishing term) is greater than 0.

International Journal of Network Security & Its Applications (IJNSA) Vol.8, No.4, July 2016

Each term in the PT set is given a weight denoted by TW, and given by: $TW = \frac{TDF_i}{N}$

Where TDFi is the term document frequency for term i in the PT data set, N is the number of phishing terms in the PT data. Table 2 depicts a sample of phishing terms and their respective document frequency and weight.

The Phishing terms weight feature for each email is the sum of the weights of the phishing terms in that email, and given by $\sum_{i=1}^{n} TW_i$, where n is the number of phishing terms in the email. The value of this feature indicates the weight of the phishing terms in that email.

Table 1: Email extracted features

| Feature | Description | Data Type | Information Gain |
|---|---|---|---|
| HTML Body | Checks if the email body contains HTML content. | Number {0,1} | 0.681 |
| Hexadecimal URLs | The number of URLs consisting of hexadecimal characters in the email. | Number | 0.652 |
| Domains Count | The number of domains in the URLs that exists in the email. | Number | 0.652 |
| TextLinkDifference | The number of URLs whose label is different from its anchor in the email. | Number | 0.649 |
| Dots Count | The maximum number of dots that exist in a URL in the email. | Number | 0.497 |
| Email Contains Account Term | Checks if the email contains the term "Account" | Number {0,1} | 0.493 |
| Email Contains Dear Term | Checks if the email contains the term "Dear" | Number {0,1} | 0.375 |
| Images as URL | The number of image URLs. | Number | 0.298 |
| IP URLs | The number of URLs whose domain is specified as an IP address. | Number | 0.297 |
| Email Contains PayPal Term | Checks if the email contains the term "PayPal" | Number {0,1} | 0.296 |
| Email Contains Login Term | Checks if the email contains the term "Login" | Number {0,1} | 0.250 |
| Email Contains Bank Term | Checks if the email contains the term "Bank" | Number {0,1} | 0.213 |
| Phishing Terms Weight | A weight that is assigned to each email and represents the sum of weights of the phishing terms that exists in that email | Number | 0.210 |
| Email Contains Verify Term | Checks if the email contains the term "Verify" | Number {0,1} | 0.207 |
| Email Contains Agree Term | Checks if the email contains the term "Agree" | Number {0,1} | 0.206 |
| Email Contains Suspend Term | Checks if the email contains the term "Suspend" | Number {0,1} | 0.205 |

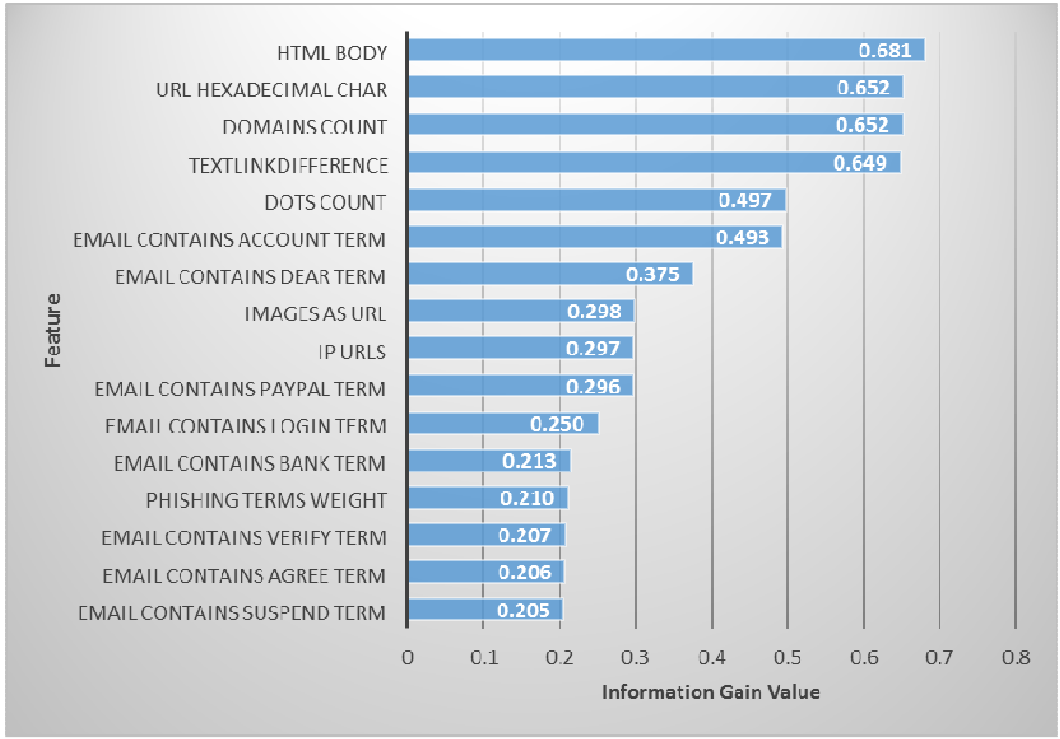

Figure 4: Features IG values

Table 2: Sample phishing terms weights

| Phishing Term | TDF | TW |
|---|---|---|
| Account | 3384 | 2.256 |
| Click | 2550 | 1.7 |
| PayPal | 1172 | 0.781 |
| Bank | 1168 | 0.779 |
| Passcode | 20 | 0.013 |

The email's phishing terms weight feature could be calculated using the following pseudocode:

```
N is the number of phishing terms in the phishing
emails corpus.
T is the set of phishing terms in the email.
 TW is an array that contains each phishing term weight.
W =0; //the phishing terms weight for the email.

For t in T loop
 W+ = TWt , where TWt is the weight of phishing term t.
End loop;
W=W/N;
```

### 3.4 Classification Model Building

After extracting the set of features from the training data set, we tested the classification accuracy of our model using five well known classification techniques; J48, Naïve Bayes, Support Vector Machine (SVM), Multi-Layer Perceptron and Random Forest. Before exploring the classification results for each algorithm, a brief summary of each algorithm's technique is presented as follows:

**J48 algorithm:** is the java implementation of the C4.5 classification algorithm, it uses a set of training data (S) consisting of already classified samples in the form $S=s_1, s_2, …, s_n$. Each sample s in the training data set consists of k-dimensional vector $(x_1, x_2, …., x_k)$, where $x_k$ represents the feature value of that sample. The algorithm constructs a decision tree from the training data set, where each node of the tree is realized by the feature that most effectively splits its set of samples into subsets using the information gain value. The main advantages of decision trees are their simplicity to explain and interpret and take into account the features relationships and interactions, however they do not support online learning and require rebuilding the tree each time new samples exists.

**Naïve Bayes Classifier:** this classifier uses the Bayes rule of conditional probability and makes use of all the data features, and analyses them individually on the assumption that they are equally important and independent of each other. The advantages of this classifier is its simplicity and quick convergence, however it cannot learn about the interactions and relationships between the features in each sample.

**SVM:** Support Vector Machine is a supervised machine learning algorithm that is mostly used for classification tasks in addition to regression tasks. In SVM each data item is plot as a point in n-dimensional space (n is the number of features in each sample in the training set) and the algorithm mission is to find the best hyper-plane that divides the two classes. SVM classifies non-linearly separable data by transforming them into a higher dimensional space (using a kernel function) where a separating hyperspace exists. SVM is known for its accuracy and its ability to classify data that is not linearly separable. However, SVM is memory-intensive and hard to interpret.

**Multi-Layer Perceptron (MLP)**: is a feed forward artificial neural network that consists multi layers (usually 3) of neurons, each neuron is considered a processing unit that is activated using an activation function. MLP is a supervised machine learning method in which the network is trained using a labelled training data set, a trained MLP will be able to map a set of input data (email features in our case) into a set of outputs (email class).

**Random Forest:** is decision tree based classification algorithm that is suitable for large data sets; it constructs a set of decision trees at training phase such that each tree operates on a predefined number of attributes chosen randomly. The classification process takes place by a majority vote of the results from each individual tree. Random Forest is trained on different parts of the training data set and aims at solving the problem of overfitting that is usually faced when using decision trees.

### 3.5 Performance metrics

In order to evaluate our proposed phishing email classification model using different classification techniques, we applied a set of evaluation metrics for each algorithm:

International Journal of Network Security & Its Applications (IJNSA) Vol.8, No.4, July 2016

- **True Positive Rate (TP):** the percentage of phishing emails in the training data set that were correctly classified by the algorithm. Formally, if the number of phishing emails in the data set is denoted by P and the number of correctly classified phishing emails by the algorithm is denoted by $N_p$, then

$$TP = \frac{N_p}{P} \quad (1)$$

- **True Negative Rate (TN):** the percentage of legitimate emails that were correctly classified as legitimate by the algorithm. If we denote the number of legitimate emails that were correctly classified as legitimate by $N_L$ and the total number of legitimate emails as L, then

$$TN = \frac{N_l}{L} \quad (2)$$

- **False Positive Rate (FP):** is the percentage of legitimate emails that were incorrectly classified by the algorithm as phishing emails. If we denote the number of legitimate emails that were incorrectly classified as phishing by $N_f$, and the total number of legitimate emails as L, then

$$FP = \frac{N_f}{L} \quad (3)$$

- **False Negative Rate (FN):** the number of phishing emails that were incorrectly classified as legitimate by the algorithm. If we donate the number of phishing emails that were classified as legitimate by the algorithm by $N_{pl}$ and the total number of phishing emails in the data set is denoted by P, then

$$FN = \frac{N_{pl}}{P} \quad (4)$$

- **Precision:** measures the exactness of the classifier; i.e. what percentage of emails that the classifier labeled as phishing are actually phishing emails, and it is given by:

$$Precision = \frac{TP}{TP+FP} \quad (5)$$

- **Recall:** measures the completeness of the classifier results; i.e. what percentage of phishing emails did the classifier label as phishing, and is given by:

$$Recall = \frac{TP}{TP+FN} \quad (6)$$

- **F-measure:** also known as F-score, and is defined as the harmonic mean of Precision and Recall, and given by:

$$F-measure = \frac{2*Precision*Recall}{Precision+Recall} \quad (7)$$

- **Receiver Operating Characteristic (ROC) Area:** a metric that demonstrates the accuracy of a binary classifier by plotting TP against FP at various threshold values.

64

International Journal of Network Security & Its Applications (IJNSA) Vol.8, No.4, July 2016

## 4. RESULTS AND DISCUSSION

This section presents the results that the proposed classification model achieved by applying the five proposed classification algorithms to the features extracted from the data set of 10538 emails including 5940 ham emails from spam assassin project [30] and 4598 spam emails from Nazario phishing corpus [31]. The generated features were fed to the five classifiers, namely J48, Bayes Net, SVM, MLP and Random Forest. To avoid overfitting, we used 10-fold cross validation technique which uses 0.9 of the training data set as data for training the algorithm and the remaining 0.1 of training data set for testing purposes, and repeat this division of the data set for training and testing for 10 times. The experiments were conducted using the open source WEKA data mining software [35].

The results were evaluated using the performance metrics discussed in the previous section. Table 3 depicts the weighted average of classification results for each of the algorithms.

The results show that our model achieves high accuracy rates in classifying phishing emails, and outperforms similar proposed classification schemes as we will explain in the next section, thanks to the proposed pre-processing phase and feature reduction and evaluation process in the proposed model. The inclusion of features with high information gain values yielded a high influence in improving the classification results. A comparison of the different algorithms results is plotted in figure 5.

The best results were achieved by the Random Forest classification algorithm, due to their usage of tree ensembles that are capable of dealing with non-linear features that are correlated to each other, and its bagging mechanism enables it to handle very well high dimensional spaces as well as large number of training examples which fits to our proposed model.

Table 3: Classification Algorithms Accuracy results (Weighted Average)

| Metrics<br>Algorithm | TP | FP | Precision | Recall | F-Measure | ROC Area |
|---|---|---|---|---|---|---|
| J48 | 0.984 | 0.019 | 0.984 | 0.984 | 0.984 | 0.9863 |
| Bayes Net | 0.954 | 0.066 | 0.947 | 0.945 | 0.945 | 0.9717 |
| SVM | 0.969 | 0.039 | 0.97 | 0.969 | 0.969 | 0.9650 |
| Random Forest | 0.991 | 0.011 | 0.991 | 0.991 | 0.991 | 0.9988 |
| MLP | 0.977 | 0.026 | 0.977 | 0.977 | 0.977 | 0.9870 |

65

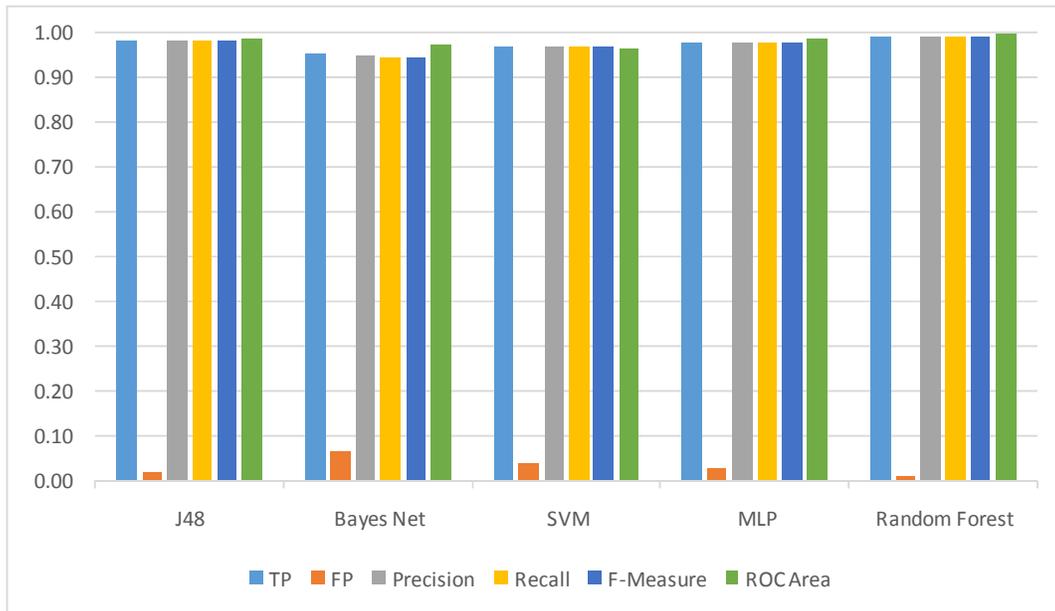

Figure 5: Classification results

The Random Forests algorithm builds a set of different decision trees for classification; to classify a new mail from the input dataset, Random Forest put the new email's features vector down each of the trees in the forest, and then a classification is obtained from each of the trees, and a classification with the most votes is returned by the algorithm. The ROC area diagram in figure 6 shows the accuracy of the random forest algorithm in separating phishing emails form legitimate ones.

We empirically evaluated the best number of trees to be used by random forest, the algorithm performed best when we set the number of trees to 30. The algorithm achieved 0.988 accuracy and 0.014 FP rate when the number of trees was set to 10. Increasing the number of trees above 30 did not add a notable improvement to the classification results.

The J48 decision tree classifier achieved the second best classification results with 0.984 TP rate and 0.019 FP rate, and yielded a small enhancement over similar studies that implemented the same algorithms for phishing email detection such as the study in [8]. The J48 algorithm achieved 0.9863 ROC area accuracy as shown in figure 7.

The third best result was achieved using the MLP classifier with TP rate of 0.977 and 0.026 FP rate. The MLP achieved a ROC area of 0.987 as shown in figure 8. SVM and Bayes Net classifiers yielded a lower percentage of classification accuracy using the proposed feature set.

International Journal of Network Security & Its Applications (IJNSA) Vol.8, No.4, July 2016

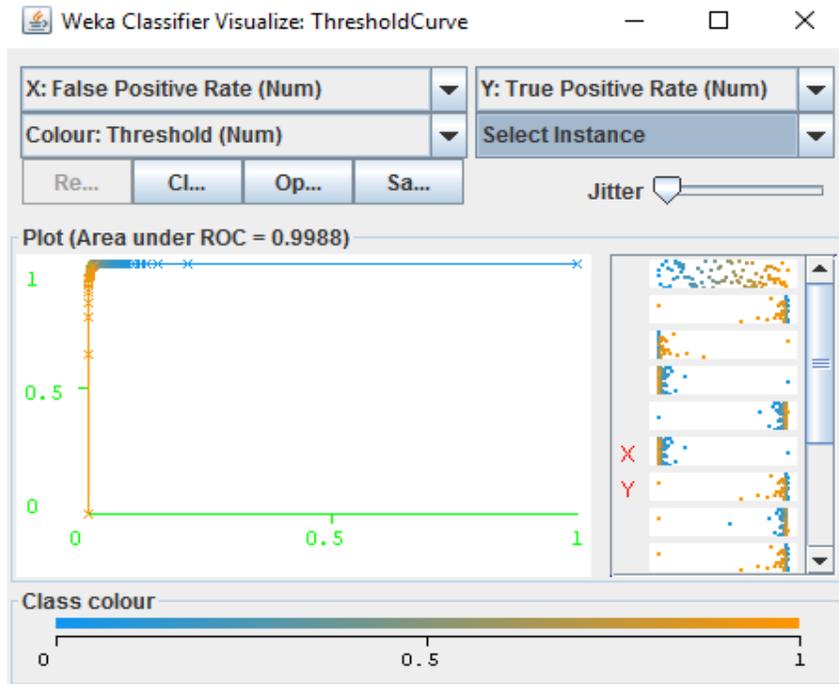

Figure 6: Random Forest ROC Area

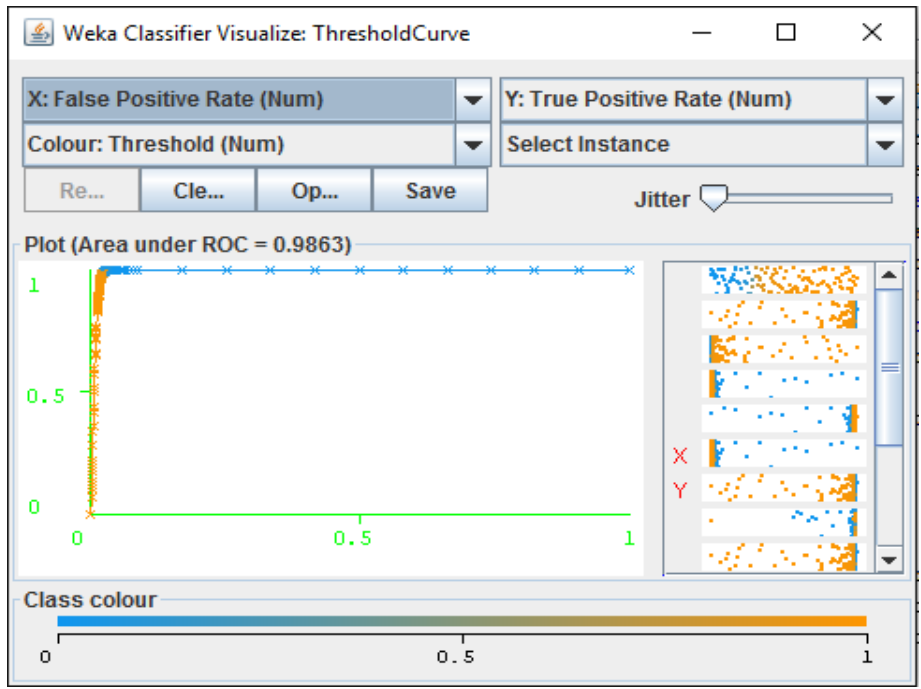

Figure 7: J48 ROC Area

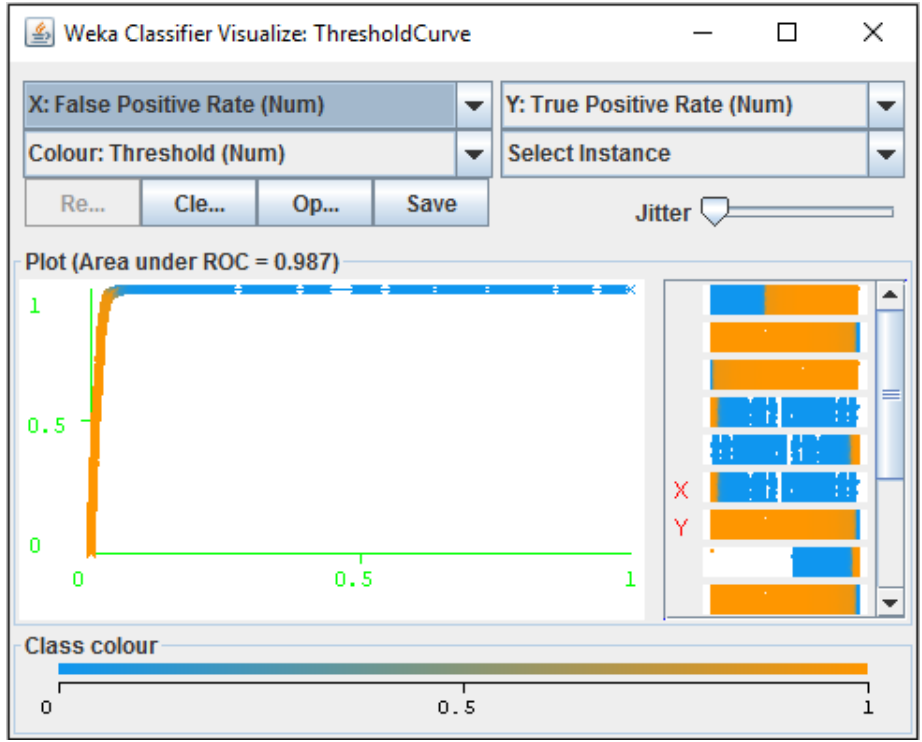

Figure 8: MLP ROC Area

## 5. COMPARATIVE ANALYSIS

A set of proposed studies are found in the literature of phishing email detection using data mining techniques, in this section we compare our proposed model with a set of previously proposed models for phishing detection. Table 4 summarizes a set of seven previous related works along with the classification algorithm(s) used and the accuracy of the classification results, the results are visualized in figure 9.

Table 4: Comparison of our approach with previous work

| Paper Reference | Classification Algorithms | Accuracy |
| --- | --- | --- |
| [36] | Random Forest | 0.97 |
| [37] | J48 + SVM | 0.97 |
| [38] | SVM | 0.75 |
| [39] | decision trees, random forest, multi-layer perceptron, Naïve Bayes and SVM | 0.99 |
| [22] | C5.0 | 0.97 |
| [40] | Bayes Net | 0.96 |
| [8] | Random Forest, LibSVM, Bayes Net, SMO, Logistic Regression and NaiveBayes. | 0.9811 |
| Our Approach | J48 , Bayes Net, SVM, Random Forest and Multi-Layer Perceptron. | 0.991 |

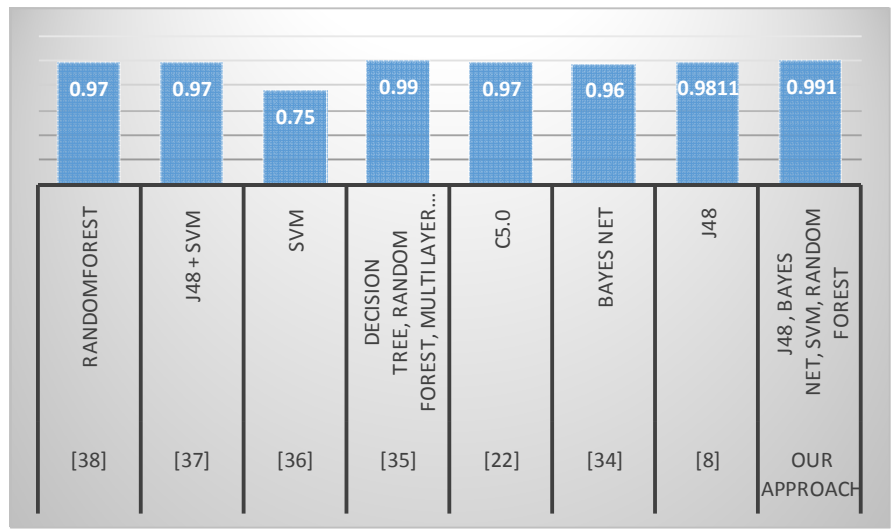

Figure 9: comparison of our approach accuracy with related work.

The study in [36] used a feature vector of 47 features extracted from the same data sets of Nazario [31]and Spam Assassin corpus [30], using Random Forest algorithm for training the classification model. Their model achieved 0.97 accuracy. Our model outperforms their model in accuracy rate with less feature set.

The study in [37] applied both J48 and SVM for classifying emails using a feature set of 30 features and yielded an accuracy rate of 0.97, our approach outperforms this result using the same classification algorithm J48 with a classification accuracy of 0.984.

The study in [38] applied the SVM algorithm only on a feature set 25 features extracted from the email content only and achieved a low accuracy rate of 0.75, our model outperforms this result due to extracting features not only from the email body, but also from the header and also using the concept of phishing terms frequency.

The study in [39] achieved high rate of accuracy in classifying phishing emails, it used a group of classification algorithms including Random Forest, Multi-Layer Perceptron, SVM and decision trees. However, this study was built on a small and not verified phishing data set.

The study in [22] achieved accuracy rate of 0.97 using the C 5 decision tree algorithm on a 22 features from two data sets of Ham, Spam emails. This result degraded to 0.84 when a third data set of Phishing emails was added.

The study in [40] achieved 0.96 accuracy using Bayes Net algorithm with seven hybrid features. However, this study was built over a small data set of 1645 emails, and when the data set was increased to 4594 emails the accuracy degraded to 0.92, and this is an indicator that their model has not been generalized.

The study in [8] achieved an accuracy rate of 0.9811 and FP rate of 0.53 using the J48 algorithm and 23 hybrid features. Our approach enhances this result to accuracy of 0.984 using less features but with FP rate of 0.019 using J48, and accuracy of 0.991 and FP rate of 0.011using Random Forest. We believe that including only features that have IG values over the data set and introducing the feature of phishing terms weight for each email contributed to this enhancement in accuracy.

## 6. CONCLUSION

This paper proposed a classification model for emails into phishing or legitimate by applying the knowledge discovery and data mining techniques, the model was built using an intelligent pre-processing phase that extracts a set of features from the emails header, body and terms frequency.

The features are enriched with WordNet ontology and text pre-processing technique of stemming to enhance the similarity between emails messages of a specific class. The extracted features were evaluated using the Information Gain measure and only those who have an information gain contribution were added to the feature set. Two accredited data sets were used in training and testing of the proposed model and 10-fold cross validation technique was used in the training and testing processes to overcome the overfitting problem. The model was experimented using five popular data mining algorithms; Random Forest, J48, SVM, MLP and Bayes Net. The classification results achieved were encouraging and enhanced the classification accuracy so far registered in similar previously published models.

As future work, the proposed model could be further enhanced by developing an adaptive mechanism to reflect the contributions of analysing new emails term frequency and applying enhanced linguistic processing techniques to strengthen the similarity between phishing emails terms such that a better classification results are obtained.

## AUTHORS

**Adwan Yasin** is an associate Professor, Former dean of Faculty of Engineering and Information Technology of the Arab American University of Jenin, Palestine. Previously he worked at Philadelphia and Zarka Private University, Jordan. He received his PhD degree from the National Technical University of Ukraine in 1996. His research interests include Computer Networks, Computer Architecture, Cryptography and Networks Security.

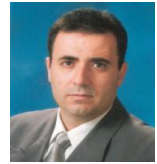

**Abdelmunem Abuhasan** is a Master student at the Arab American University with particular interests in computer security, web security and software engineering. He is working since ten years as the manager of software development department at the Arab American University. He holds a B.A. in Computer Science from the Arab American University.

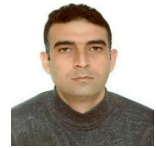